\documentclass[aps,10pt,amsmath,amssymb]{revtex4}
\usepackage{amsmath} \usepackage{subfigure} \usepackage{graphicx}
\usepackage{rotating}
\usepackage{multirow}

\newcommand{\fes}{Fe$_2$S$_2$ } 
\newcommand{\Cr}{Cr$_2$ } 

\begin{document}

\title{A spin-adapted Density Matrix Renormalization Group algorithm for quantum chemistry} 
\author{Sandeep
  Sharma}
\author{Garnet Kin-Lic Chan\footnote[1]{Corresponding author. Electronic mail: gc238@cornell.edu}}
\affiliation{Department of Chemistry and Chemical Biology, Cornell University, Ithaca NY14853}

\begin{abstract}
We extend the spin-adapted density matrix renormalization group (DMRG) algorithm of McCulloch and Gulacsi \cite{mcculloch3}
to quantum chemical Hamiltonians.
This involves two key modifications to the non-spin-adapted DMRG algorithm:
the use of a quasi-density matrix to ensure that the renormalised DMRG states are eigenvalues of $\hat{S}^2$,
and the use of the Wigner-Eckart theorem  to greatly reduce the overall storage and computational
cost. We argue that the advantages of the spin-adapted DMRG algorithm are greatest for low spin states.
Consequently, we also implement the singlet-embedding strategy
 of Nishino \textit{et al} \cite{tatsuaki} which allows us to target high spin states as a component
of a mixed system which is overall held in a singlet state. We evaluate our algorithm on benchmark calculations on
the {\fes} and {\Cr} transition metal systems. By calculating the full spin ladder of {\fes}, we  show that the
spin-adapted DMRG algorithm can
target very closely spaced spin states. In addition, our calculations of {\Cr}
demonstrate that the spin-adapted algorithm requires only roughly half
the number of renormalised DMRG states as the non-spin-adapted algorithm to obtain the same accuracy in the energy,
thus yielding up to an order of magnitude increase in computational efficiency.
\end{abstract}

\maketitle

\section{Introduction}

Since its introduction by White~\cite{White1992,White1993} and its first application to 
quantum chemical systems~\cite{White1999}, the density matrix renormalization group (DMRG) has been applied to 
a wide variety of problems in quantum chemistry~\cite{Mitrushenkov2001,Chan2002,Marti2008,Legeza2003dyn,Zgid2008spin,kurashige}. 
After  early attempts to use the DMRG as a full configuration interaction (FCI) method for
small molecules~\cite{Daul2000,Chan2002, Mitrushenkov2003,Zgid2008spin,Legeza2003lif},
it was recognised that DMRG is best used to describe
non-dynamical correlation in active spaces. The DMRG algorithm exhibits a polynomial cost scaling $O(k^3 M^3) + O(k^4 M^2)$, 
where $k$ is the number of active space orbitals, and $M$ is the number of renormalised many-body states which
determine the accuracy of the method. In non 1-D systems, the number of states $M$ required to obtain
a given error (relative to the FCI energy in the active space) depends on
the correlation length of the system with the orbitals mapped onto an artificial 1-D lattice,
and this can increase quite rapidly with $k$.
In addition, the shape of the orbitals and the order in which they are
arranged can drastically affect the convergence of the DMRG~\cite{Rissler2006,Legeza2003qie}. 
Nonetheless, many examples have demonstrated that in practical applications, the DMRG
 describes active space correlations to high accuracy, for orbital spaces beyond the reach of
complete active space non-dynamical correlation methods.

Transition metal chemistry typically involves partially filled $d$ orbitals 
and is a rich source of difficult active space correlation problems. 
Increasing effort in recent times has been devoted to applications of the DMRG to 
transition metal chemistry\cite{yanaiCT,Moritz2005orb,Moritz2006,Moritz2005rel,Moritz2007,Marti2008, kurashige}.
Here, the ability to correctly handle spin symmetry is an important asset.
This is because the large number of unpaired electrons often leads to 
many low lying spin states in a very narrow energy window. These can only be efficiently resolved by targetting a specific spin sector.
In addition, of course, the correct use of spin symmetry offers the promise of  computational
efficiency gains.

Spin symmetry is associated with the non-Abelian SU(2) Lie group. Spin adaptation in the DMRG can be achieved
by working with states and operators
(multiplets and irreducible tensor operators, respectively) that transform as irreducible
representations of SU(2). This formulation resembles quantum chemistry approaches to spin adaptation which work directly in the
 configuration state function basis, rather than alternatives based on the symmetric\cite{Duch,Rud} or unitary groups~\cite{wormer,shavitt,brooks}.
The first DMRG algorithm  to exploit non-Abelian spin symmetry was the interaction-round-a-face DMRG (IRF-DMRG) 
introduced by Sierra \textit{et al.}\cite{sierra, wada}. McCulloch \textit{et al.}~\cite{mcculloch1,mcculloch2,mcculloch3} later 
proposed a highly efficient implementation of spin-adapted DMRG. Their algorithm relied on two important ingredients. 
The first was the use of a 
quasi-density matrix to determine the renormalized DMRG basis.
In general, the density matrix of a subsystem does not commute with the total spin operator of the subsystem,
and thus the usual DMRG prescription, to use the density matrix eigenvectors as the many-body basis,
 is incompatible with spin adaptation. McCulloch \textit{et al.} showed that the best states to retain in the decimation step 
of the DMRG are eigenvectors of a \textit{quasi-density} matrix which commutes with the $\hat{S}^2$ operator. The second
 contribution  was the use of the Wigner-Eckart theorem  to efficiently store and compute
matrix elements of irreducible tensor operators. This leads to significant improvements in the performance of DMRG. 
In this work, we closely follow McCulloch \textit{et al.} and extend their algorithm to deal with the 
more complicated Hamiltonians in quantum chemical systems. We note that earlier work on spin-adapted
DMRG in the context of quantum chemistry was carried out by Zgid \textit{et al.}~\cite{Zgid2008spin}. Zgid \textit{et al.} used 
quasi-density matrices to ensure the proper spin symmetry of the renormalised states but did not take advantage
 of the Wigner-Eckart theorem. As we will show, while the Wigner-Eckart formulation complicates
the implementation of the DMRG algorithm significantly, it also results in substantial performance gains.

We start with a brief summary of the DMRG algorithm in Section~\ref{sec:dmrg_algorithm}. We assume that the reader has
some familiarity with  the DMRG algorithm as described
 in
various articles~\cite{Chan2002,Schollwock2005,kurashige,chan2011}, thus we focus mainly on aspects of the DMRG 
that will be modified when spin adaptation is introduced. In section~\ref{sec:spin} we describe in some detail our implementation of spin adaptation in DMRG. We review concepts related to spin symmetry, such as the Wigner-Eckart theorem, Clebsch-Gordan coefficients, 6-\textit{j} coefficients, and 9-\textit{j} coefficients, although the reader will benefit from more detailed
expositions, for example in Refs.~\cite{edmonds,brink}. In section~\ref{sec:computational}  we present our
 analysis of the main computational differences between the spin-adapted and non-spin-adapted algorithms and describe the singlet embedding approach to high spin states. Finally in Section~\ref{sec:applications} we present some sample calculations on transition metal systems, that demonstrate 
the advantages of using the spin-adapted DMRG algorithm. The appendices summarise  some useful relations between the 
various Clebsch-Gordan coefficients, and describes spin adaptation in the matrix product state language.


\section{A summary of the DMRG algorithm}

\label{sec:dmrg_algorithm}
\begin{figure}
\begin{center}
\resizebox{80mm}{!}{\includegraphics{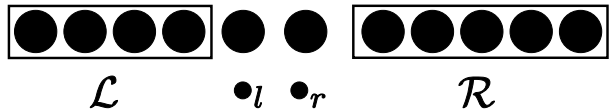}}
\end{center}
\caption{The one-dimensional arrangement of orbitals on a lattice and
  the subdivision into blocks. In the forward sweep the left block is
  termed the system block and the right block is termed the
  environment block and the reverse is true in the backward sweep. At each sweep iteration the system block
  increases in size by one orbital.}
\label{fig:dmrg}
\end{figure}

The basic DMRG algorithm consists of a set of sweeps over the $k$ spatial orbitals
of the problem. We imagine these orbitals to be arranged as a one
dimensional lattice of sites. At every step of the algorithm, the lattice is conceptually divided into four parts: a
{left block} $\mathcal{L}$ consisting of sites  $1 \ldots p-1$,
a {left dot} $\bullet_l$, consisting of site $p$, a {right dot}  $\bullet_r$
consisting of site $p+1$, and a {right block} $\mathcal{R}$ consisting of
sites $p+2 \ldots k$ (see Figure~\ref{fig:dmrg}). In the forward sweeps, the
orbital index $p$ increases from $2 \ldots k-2$,  and block $\mathcal{L}$ increases
in size to cover the lattice, while block $\mathcal{R}$ shrinks.
 During the backwards sweeps, the index $p$  iterates backwards from $k-2 \ldots 2$, and block $\mathcal{R}$ increases
in size to cover the lattice, while block $\mathcal{L}$ shrinks.
When it is necessary to refer to blocks at different sweep iterations, we will use additional subscripts to indicate the sites spanned by block. For example, in successive iterations in a forward sweep, the two $\mathcal{L}$ blocks would be $\mathcal{L}_{p-1}$ (sites $1 \ldots p-1$) and $\mathcal{L}_p$ (sites $1 \ldots p$), and the two left dots would be $\bullet_p$ and $\bullet_{p+1}$. We refer to the set of computations performed at each value of index $p$ as a {sweep iteration};
 a sweep thus contains $k-4$ sweep iterations. In total, the full calculation consists of multiple forwards and backwards
sweeps (each containing multiple sweep iterations) until convergence in the energy is observed.

Blocks $\mathcal{L}$ and $\mathcal{R}$ are each associated with $M$ 
many body states, denoted by $\{|l\rangle\}$ and
$\{|r\rangle\}$ respectively, where the state labels range from $l, r=1\ldots M$. (If we need to be more specific about the nature of the block we will attach subscripts e.g. block $\mathcal{L}_{p-1}$ contains states ${|l_{p-1}\rangle}$.) In successive sweeps of the DMRG algorithm,
these many body spaces are variationally improved. The left and right dots are associated
with the complete Fock spaces of their respective orbitals $\{ |n_l\rangle\}$,
$\{|n_r\rangle\}$ respectively, where $ |n\rangle \in \{|-\rangle, |\alpha\rangle, |\beta\rangle, |\alpha
\beta \rangle\}$.

During the calculation we wish to calculate observables, that is,  expectation values of  operators such as the Hamiltonian.
In general such operators can be expressed as (sums of) products of operators partitioned between the four blocks. 
For example, a two particle density matrix element operator
$a^\dag_i a^\dag_j a_k a_l$ is partitioned amongst the blocks depending on the values of the indices $i, j, k, l$. (Note we use the indices to specify spin orbitals; later while describing the spin-adapted algorithm the indices will be used to specify spatial orbitals. The distinction will be clear from the context). 
The Hamiltonian  across the whole lattice
involves sums of the density matrix element operators, and can thus be partitioned in multiple ways
into operators on each of the different blocks.
\begin{align}
 \hat{H} =& \sum_{ij} t_{ij} a_i^\dag a_j + \frac{1}{2}\sum_{ijkl 
} v_{ijlk} a_i^\dag a_j^\dag a_k a_l 
\end{align}
The following set of operators and their adjoints, defined in Table \ref{tab:abelian_ops}, provides an efficient partitioning:
$\hat{1}, a_i, \hat{A}_{ij}, \hat{B}_{ij}, 
\hat{R}_i, \hat{P}_{ij}, \hat{Q}_{ij}, \hat{H}$~\cite{xiang}.
$\hat{R}_i, \hat{P}_{ij}, \hat{Q}_{ij}$ are known as complementary operators, and their definitions involve the one- and two-electron integrals.

\begin{table}
 \begin{center}
\caption{Definition of the operators used in the DMRG algorithm. Here the indices are spin
  indices not spatial indices.}
  \begin{tabular}{ll}
   \hline \hline
Operator& Definition\\ 
\hline 
$\hat{A}_{ij}$ & $a_i^\dag a_j^\dag$\\
$\hat{B}_{ij}$ & $a_i^\dag a_j$ \\
$\hat{R}_i$ & $\sum_i t_{ij} a_j + \sum_{jkl} v_{ijlk} a_j^\dag a_k a_l$ \\
$\hat{P}_{ij}$ & $\sum_{kl} v_{ijlk} a_k a_l$ \\
$\hat{Q}_{ij}$ & $\sum_{kl} (v_{ikjl} - v_{iklj}) a_k^{\dag} a_l$ \\
\hline
\hline
  \end{tabular}
 \end{center}
\label{tab:abelian_ops}
\end{table}

The computations in a sweep iteration consists of manipulations of
states and operators in the  spaces associated with the four blocks $\mathcal{L}, \bullet_{l}, \bullet_{r}, \mathcal{R}$.
These computations are divided into three steps
blocking, wavefunction solution, and renormalization and decimation. We now
describe these computations in the context of a forward sweep.

\textit{Blocking ---} This consists, conceptually, of adding the left dot to the left
  block and the right dot to the right block to form blocks $\mathcal{A}=\mathcal{L}\bullet_{l}$ and
  $\mathcal{B}=\bullet_{r} \mathcal{R}$, respectively. Blocks $\mathcal{A}$ and $\mathcal{B}$ are each associated
with  many body
  spaces $\{ |a\rangle \}$, $\{ |b\rangle \}$, where the state labels range from $a, b=1\ldots 4M$. They are  product
    spaces i.e.  $\{|a\rangle \} = \{ |l\rangle\} \otimes \{|n_l\rangle \}$ and
     $\{ |b \rangle \} = \{|n_r\rangle\} \otimes \{|r\rangle \}$. 

During blocking, the matrix elements of operators on block $\mathcal{A}$ and block $\mathcal{B}$
are formed from the matrix elements of constituent operators on the blocks $\mathcal{L}$,  $\bullet_{l}$ and $\bullet_{r}$, $\mathcal{R}$ respectively. Consider the operations to form the matrix representation of $\hat{A}_{ij} = a_i^{\dag} a_j^{\dag}$ on block $\mathcal{A}$. We write this as 
$\mathbf{A}_{ij}[\mathcal{A}]$, where the bold font denotes matrix representation.
Depending on the indices $i, j$, the matrix representation 
$(\mathbf{A}_{ij}[\mathcal{A}])_{aa'}= \langle a|a_i^{\dag} a_j^{\dag}|a' \rangle$ is formed in one of three ways,
 \begin{align}
 i, j \in \mathcal{L} &\Rightarrow \mathbf{A}_{ij}[\mathcal{L}] \otimes \mathbf{1}[\bullet_l] \notag \\
 i \in \mathcal{L}, j \in \bullet_l &\Rightarrow \mathbf{a}_i[\mathcal{L}] \otimes \mathbf{a}_j[\bullet_l] \notag \\
 i, j \in \bullet_l &\Rightarrow \mathbf{1}[\mathcal{L}] \otimes \mathbf{A}_{ij}[\bullet_l] 
 \label{eq:aiaj_abblocking}
 \end{align}
Here $\otimes$ denotes a tensor product between operators that is defined with a parity factor to take into account fermion statistics. For
two  operators $\hat{X}$ and $\hat{Y}$ with matrix elements  $\langle \mu | \hat{X} | \mu'\rangle$, 
$\langle \nu | \hat{Y} | \nu'\rangle$, the tensor product is defined through
\begin{align}
\langle \mu \nu | \hat{X} \hat{Y} | \nu' \mu' \rangle = \mathcal{P}(\nu,\hat{X}) \langle \mu | \hat{X} | \mu' \rangle \langle \nu | \hat{Y} | \nu' \rangle \label{eq:otimes}
\end{align}
where $\mathcal{P}$ is the fermionic parity operator.
Similarly, the Hamiltonian matrix $\mathbf{H}[\mathcal{A}]$ is built from the matrix representations of operators
in Table \ref{tab:abelian_ops}  acting on blocks $\mathcal{L}$, $\bullet_l$,
\begin{align}
\mathbf{H}[\mathcal{A}] =& \mathbf{H}[\mathcal{L}] \otimes \mathbf{1}[\bullet_l] + \mathbf{1}[\mathcal{L}] \otimes \mathbf{H}[\bullet_l]  \nonumber\\
&+ \frac{1}{2}\sum_{i\in \mathcal{L}} \left(\mathbf{a}_i^{\dag}[\mathcal{L}] \otimes\mathbf{R}_i[\bullet_l] + \mathbf{R}_i^{\dag}[\bullet_l] \otimes\mathbf{a}_i[\mathcal{L}]\right) \nonumber\\
&+ \frac{1}{2}\sum_{i\in \bullet_l} \left(\mathbf{a}_i^{\dag}[\bullet_l]\otimes \mathbf{R}_i[\mathcal{L}] + \mathbf{R}_i^{\dag}[\mathcal{L}]\otimes \mathbf{a}_i[\bullet_l]\right) \nonumber\\
&+ \frac{1}{2}\sum_{ij\in \bullet_l} \left(\mathbf{A}_{ij}[\bullet_l]\otimes \mathbf{P}_{ij}[\mathcal{L}] +   \mathbf{A}_{ij}^{\dag}[\bullet_l]\otimes\mathbf{P}_{ij}^{\dag}[\mathcal{L}]\right)\nonumber\\
&+ \frac{1}{2}\sum_{ij\in \bullet_l} \mathbf{B}_{ij}[\bullet_l] \otimes\mathbf{Q}_{ij}[\mathcal{L}]
\label{eq:hblocking}
\end{align}
The representation of other operators in Table \ref{tab:abelian_ops} for block $\mathcal{A}$ may be constructed by formulae analogous to Eqs. (\ref{eq:aiaj_abblocking}) and (\ref{eq:hblocking}).
These formulae are summarised in  Appendix \ref{sec:appendix_blocking}.

\textit{Wavefunction solution ---} Here we solve for a target eigenstate of $\hat{H}$ for the full problem of  $k$ orbitals.
In DMRG the corresponding Hilbert space 
is spanned  by the product basis of  $\mathcal{A}$ and $\mathcal{B}$, which we refer to as the superblock space $\{ |ab\rangle \}$. 
The corresponding matrix representation of  $\hat{H}$ is the superblock Hamiltonian $\mathbf{H}[\mathcal{AB}]$.
The superblock Hamiltonian $\mathbf{H}[\mathcal{AB}]$ is (formally) defined from Eq. (\ref{eq:hblocking}), where 
$\mathcal{A}$, $\mathcal{B}$ replace the block labels $\mathcal{L}$, ${\bullet}_l$. 
Note that we could also rewrite the Hamiltonian formula in Eq. (\ref{eq:hblocking}) with the labels $\mathcal{A}$ and $\mathcal{B}$ swapped.
For  efficiency, we use the above definition when the number of orbitals in block $\mathcal{A}$ is larger
than that in block $\mathcal{B}$, and swap the labels $\mathcal{A}$ and $\mathcal{B}$ when the reverse is true.

The superblock Hamiltonian matrix is never built in practice, as we only
wish to obtain one (or a few) eigenvectors. Instead the target wavefunction is expanded in the superblock basis $\{|ab\rangle\}$
\begin{align}
|\Psi\rangle = \sum_{ab} \mathbf{C}_{ab} |ab\rangle = \sum_{ln_ln_rr}\mathbf{C}_{ln_ln_rr} |ln_ln_rr\rangle\label{eq:wfn}
\end{align}
and we obtain the  eigenvector $\mathbf{C}$ using
 the Davidson algorithm.
The main operation in the Davidson algorithm is the Hamiltonian wavefunction product $\mathbf{H} \cdot \mathbf{C}$.
Since $\mathbf{H}$ is partitioned into a sum of products of operators on blocks $\mathcal{A}$ and $\mathcal{B}$ as  
Eq. (\ref{eq:hblocking}), this is carried out for each term in the sum, defining suitable intermediates. For example, 
\begin{align}
(\mathbf{A}_{ij}[\mathcal{A}]
\otimes \mathbf{P}_{ij}[\mathcal{B}]) \cdot \mathbf{C} 
= \mathbf{A}_{ij}[\mathcal{A}] \mathbf{C}  \mathbf{P}_{ij}^{T}[\mathcal{B}] \label{eq:hc_intermediates}
\end{align}
and product is efficiently carried out by grouping the terms $(\mathbf{A}_{ij}[\mathcal{A}] \mathbf{C})  \mathbf{P}_{ij}^{T}[\mathcal{B}]$
or $\mathbf{A}_{ij}[\mathcal{A}] (\mathbf{C}  \mathbf{P}_{ij}^{T}[\mathcal{B}])$, where superscript $T$ corresponds to the transpose of the operator.

\textit{Renormalization and decimation ---} Here the many-body space  of block
  $\mathcal{A}$ is truncated from dimension $4M$ to dimension $M$, to obtain the
states and operators of the next $\mathcal{L}$ block
in the sweep.  As argued by White~\cite{White1992}, the optimal
  truncated space is formed by the eigenvectors of the density matrix of  $\mathcal{A}$
 with the largest eigenvalues. The density matrix is defined by tracing out the contributions of the
right block $\mathcal{B}$ to the full density matrix,
\begin{align}
\hat{\Gamma} =& \mathrm{Tr}_B|\Psi\rangle\langle \Psi| \\ 
\mathbf{\Gamma} =&  \mathbf{C} \mathbf{C}^{\dag}
 \label{eq:dm_defn}
\end{align}
The eigenvectors are obtained from
\begin{align}
\hat{\Gamma} |l\rangle &= \sigma_l |l\rangle 
\end{align}
 and the $M$ largest
  eigenvalues yield a set of eigenstates $\{ |l\rangle \}$, $l=1 \ldots M$.
We can collect  the eigenvectors into a transformation matrix $\mathbf{L}$, where
\begin{align}
\mathbf{\Gamma} \mathbf{L} = \mathbf{L} \mathrm{~diag}[\sigma_1, \ldots, \sigma_M].\label{eq:lcoeffs}
\end{align}
The remaining eigenvalues of the discarded eigenstates, $\sigma_{M+1}\ldots\sigma_{4M}$ may be summed to give a total discarded weight, which measures the accuracy of the DMRG truncation and which can be used in DMRG extrapolation to the $M=\infty$ limit.
To complete the renormalization, we need to convert block $\mathcal{A}$ into a new
left block $\mathcal{L}$. To do this, we truncate the basis $\{ |a\rangle \}$
to the renormalised space $\{ |l\rangle\}$ of dimension $M$ as above. We next project
all the operators constructed on $\mathcal{A}$ into this renormalised space.
The projection is written in terms of the density matrix eigenvectors.
For an operator $\mathbf{X}[\mathcal{A}]$, we have,
\begin{align}
\mathbf{X}[\mathcal{L}] = \mathbf{L}^\dag \mathbf{X}[\mathcal{A}] \mathbf{L}
\label{eq:optran}
\end{align}
At the end of the decimation step, we have constructed both the space and the operators
of the new block $\mathcal{L}$, and we can 
 proceed to the next sweep iteration. 

For efficient calculations, an additional operation is performed after renormalisation and decimation.
The convergence of the Davidson algorithm is greatly
improved with a good initial guess for the  coefficients $\mathbf{C}$. 
We can transform the converged coefficients $\mathbf{C}$ obtained during one step of the sweep, to
obtain a guess $\mathbf{G}$ for the wavefunction at the next step. This wavefunction transformation
uses the forward transformation matrix $\mathbf{L}$ obtained above (for block $\mathcal{L}_p$), as well as the backward transformation matrix  $\mathbf{R}$ (for block $\mathcal{R}_{p+1}$)
obtained from a backwards sweep. 


The guess wavefunction is then constructed as
\begin{align}
 \mathbf{G}_{l_{p} n_{p+1}, n_{p+2}r_{p+3}} =&\sum_{\substack{l_{p-1}n_p \\ r_{p+1}}}  \tilde{\mathbf{L}}_{l_p,l_{p-1} n_{p}} \mathbf{R}_{n_{p+2} r_{p+3},r_{p+2}} \mathbf{C}_{l_{p-1} n_{p}, n_{p+1}r_{p+2}} \label{eq:wave_transform}
\end{align}
where $\tilde{\mathbf{L}}$ is the pseudo-inverse of $\mathbf{L}$.

\subsection{Abelian symmetries in the DMRG}

 Abelian symmetries, which include, for example, the axial spin component
 $m$, total particle number $N$, and Abelian point
group symmetry, are taken into account in a straightforward manner in the DMRG.
We label each block basis state $|\mu\rangle$ by  an additional set of
quantum numbers $q$ corresponding to the irreducible representations of
all the applicable symmetries, i.e.
\begin{align}
|\mu\rangle \to |\mu q\rangle
\end{align}
For a product state, such as formed in the blocking step,
Abelian symmetry means that the quantum numbers of the product state
are just the ``sum'' of quantum numbers of the individual states
\begin{align}
|\mu q\rangle & = |\mu_1 q_1 \mu_2q_2\rangle \nonumber\\ 
q &= q_1 \oplus q_2 \label{eq:qnumber_sum}
\end{align}
In the case of $N$ and $m$, $\oplus$ is given by standard addition
(i.e. $N=N_1+N_2$) while in the case of point groups, it is given by modulo addition.

The target eigenstate obtained from DMRG transforms according to a desired
irreducible representation. Consequently
only many body states $|a\rangle$ and $|b\rangle$ whose quantum
numbers sum to the target state quantum numbers need appear in the
wavefunction expansion, 
\begin{align}
|\Psi_q\rangle &= \sum_{ab} \mathbf{C}_{aq_a bq_b} |aq_a  bq_b\rangle 
\nonumber\\ q &= q_a \oplus q_b
\end{align}
 and thus Abelian symmetry can significantly reduce the number
of coefficients in $\mathbf{C}$.

Operators on the blocks can also be labelled by Abelian symmetry representations or quantum numbers.
 For example,
$a^\dag_{i\beta}$ is labelled by particle quantum number 1 and $m$ quantum
number $-1/2$, reflecting how the operator changes the quantum numbers of the
states that it acts on. The labelling of operators by 
quantum numbers allows the use of selection rules to store and
manipulate only the non-zero elements of the operators. These take the form
\begin{align}
 \langle \mu_1q_1| \hat{X}^q| \mu_2q_2\rangle =
 \delta_{q_1,q\oplus q_{2}} \langle \mu_1 q_1| \hat{X}^{q}|
 \mu_2 q_2\rangle \label{eq:sel}
\end{align}

Labelling states and operators using Abelian symmetry thus leads to the following computational advantages:
 it reduces the number of states that need to be considered  on each block, since they need to combine to yield the
  quantum numbers of the target wavefunction, it limits the coefficients $\mathbf{C}$ in the
  wavefunction expansion, and, selection rules allow us to work with only non-zero elements
of the operators.

\section{Spin Adaptation of the DMRG algorithm}\label{sec:spin}

As discussed in the introduction, the incorporation of spin symmetry can potentially yield significant computational 
advantages in the DMRG algorithm.
The basic advantages are similar to those for Abelian symmetries: elimination
of  block states  which cannot participate in the final target wavefunction,
restriction of coefficients in the wavefunction expansion, and
selection rules to work with only the non-zero
operator elements. However, the non-Abelian nature of the SU(2) Lie group brings
additional features into play. For example, associated with every spin state $S$ is
a $2S+1$ degenerate manifold of multiplet states, but if we are interested
in the expectation value of a rotationally invariant operator such
as the Hamiltonian, then we can work with multiplets as a single entity, rather
than working with the individual states. The target wavefunction is then expanded in terms of a set of \textit{reduced coefficients}
labelled by multiplets, rather than states. Similarly
operators are represented by \textit{reduced matrix elements}, labelled  by multiplets rather than states. 
For a given particle number $N$ in an orbital space of size $k$, the relative dimension of the 
number of multiplets of spin $S$ versus the dimension of the state space with axial spin $m=S$ is given
by the ratio of the Weyl formula for the number of configuration state functions (with $m=S$)
and the formulae for the number of determinants, namely
\begin{align}
  \text{no. CSF} =& \frac{2S+1}{k+1}\left(\begin{array}{c} k+1\\ n/2-S \end{array}\right)\left(\begin{array}{c} k+1\\ n/2+S+1 \end{array}\right) \nonumber\\
  \text{no. dets}&= \left(\begin{array}{c} k\\ n/2+m \end{array}\right)\left(\begin{array}{c} k\\ n/2-m \end{array}\right) 
\end{align}

\begin{figure}
\begin{center}
\resizebox{80mm}{!}{\includegraphics{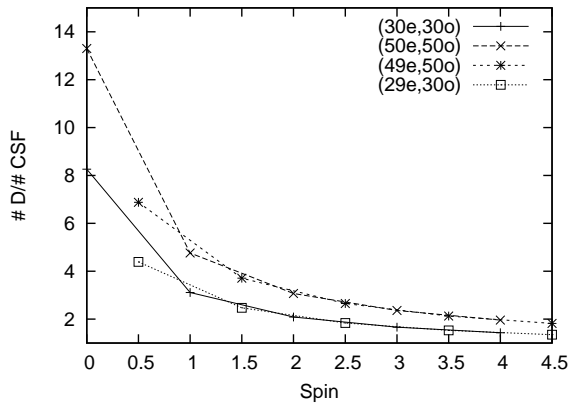}}
\end{center}
\caption{The figure shows the ratio of the the number of determinants to the number of configuration state functions of a given $m$ and $S$ respectively that can be produced with an active space shown in the legend.}
\label{fig:SS}
\end{figure}

The computational advantage of using the multiplet space, versus
the state space, is therefore a function of the particle number, number of orbitals, and spin.
Some typical ratios are shown in Fig. \ref{fig:SS}.
We see that the number of multiplets can be much smaller than the number of states, and thus the computational advantages of
using the reduced representations can be substantial, particularly when $S$ is small.

Of course, working with the reduced multiplet representations introduces some complications which involve the algebra of  SU(2).
We now recap the theory of spin eigenstates and spin tensor operators as relevant to the DMRG,
before describing the application to the steps of the sweep iteration.

\subsection{Spin eigenstates}

Spin symmetry introduce two additional quantum numbers, $S$ and $m$
\begin{align}
|\mu\rangle \to |\mu Sm\rangle.
\end{align}
Each $S$ is associated with a degenerate multiplet of $2S+1$ $m$ states, which transform
amongst each other under rotation. The non-Abelian character
 of spin is apparent when we construct  spin eigenstates
from two underlying spins. 
In this case $| Sm\rangle$ is not the product
of spin eigenstates $| S_1m_1 S_2 m_2\rangle$, but
instead a linear combination of product states with different $m_1$ and $m_2$, coupled
by  Clebsch-Gordan coefficients $c^{SS_1 S_2}_{m m_1 m_2}$,
\begin{align}
| Sm\rangle & =  \sum_{m_1 m_2} c^{S S_1 S_2}_{m m_1
m_2} | S_1 m_1  S_2 m_2\rangle \nonumber\\
m&=m_1+m_2\label{eq:spin_sum}\\
S & \in \{|S_1 - S_2|, |S_1-S_2|+1, \ldots (S_1 +S_2)\} \label{eq:s_restriction} 
\end{align}
Eq. (\ref{eq:s_restriction})  generalizes Eq. (\ref{eq:qnumber_sum})
for  Abelian symmetry, to spin symmetry.
Because of the restriction in the range of allowed $S_1, m_1, S_2, m_2$ from Eqs. (\ref{eq:spin_sum}) and (\ref{eq:s_restriction}),
we observe that spin confers a similar advantage to an Abelian symmetry
in a DMRG calculation:
block  states on $\mathcal{A}$, $\mathcal{B}$ need not be considered if they cannot
 combine to yield the $S, m$ quantum numbers in the target wavefunction. 

As mentioned above when solving the Schr\"odinger equation with spin
symmetry we can work with multiplets as a single entity,  rather than individual states, because $\hat{H}$
is invariant under rotation. \textit{Reduced quantities} are labelled only  by $S$, and the \textit{reduced} wavefunction is written as
\begin{align}
||\Psi_{S} \rangle = \sum_{a S_a b S_b} \mathbf{C}_{a S_a b S_b} ||a S_a b S_b\rangle \label{eq:gstate}
\end{align}
 The reduced coefficients in the multiplet representation are related to the coefficients $\mathbf{C}_{a S_a
  m_a b S_b m_b}$ in the state representation,
\begin{align}
|\Psi_{S m} \rangle = \sum_{a S_a m_a b S_b m_b} \mathbf{C}_{a S_a
  m_a b S_b m_b} |a S_a m_a b S_b m_b\rangle
\end{align}
by, 
\begin{align}
\mathbf{C}_{a S_a m_a b S_b m_b} &=  c^{S_a S_b S}_{m_a m_bm} \mathbf{C}_{a S_a b S_b} \label{eq:reduced_wf}
\end{align}

The reduced coefficients $\mathbf{C}_{a S_a bS_b }$ are clearly
 smaller in number than the original set of wavefunction coefficients $\mathbf{C}_{a S_a m_a b S_b m_b}$.

\subsection{Spin tensor operators}

With spin, symmetry operators can also acquire labels $S, m$. 
Operators which transform according to irreducible spin representations are known
as irreducible (spin) tensor operators.
 Similarly to a spin multiplet,  tensor operators labelled by $S$ are associated with
a manifold of $2S+1$ operators that transform amongst each other under rotation.
A simple way to characterize a tensor operator is to observe its effect on 
 a state with  spin $S=0$. For example, 
 $a^\dag_{i\alpha}$ and $a^\dag_{i\beta}$ are
2 components of a $S=\frac{1}{2}$ (doublet) tensor operator $a^{1/2}$, because they act on a
vacuum state (with spin $S=0$) to   generate  eigenstates of spin $\frac{1}{2}$. Considering
 the operators $a^\dag_{i\alpha} a_{j\alpha}, a^\dag_{i\alpha}
a_{j\beta}, a^\dag_{i\beta} a_{j\alpha}, a^\dag_{i\beta} a_{j\beta}$,
they collectively span an $S=0$ singlet and an $S=1$ triplet manifold. The
$S=0$ singlet operator is defined as
\begin{align}
\hat{B}_{ij}^{0,0}=\frac{1}{\sqrt{2}}(a^\dag_{i\alpha}a_{j\alpha} + a^\dag_{i\beta}a_{j\beta})
\end{align}
and the $S=1$ triplet  operators are defined as
\begin{align}
\hat{B}_{ij}^{1,-1}&=a^\dag_{i\beta}a_{j\alpha} \\
\hat{B}_{ij}^{1,0}&={\frac{1}{\sqrt{2}}}(a^\dag_{i\alpha}a_{j\alpha} -  a^\dag_{j\alpha}a_{i\alpha})\\
\hat{B}_{ij}^{1,1}&=-a^\dag_{i\alpha}a_{j\beta}
\end{align}
A full list of the tensor operators used in the spin-adapted DMRG algorithm is given in Table \ref{tab:ops}.

Tensor operators allow us to work with  reduced operator matrix elements, labelled only by multiplets
\begin{align}
\mathbf{X}^S_{\mu_1 S_1 \mu_2 S_2} = \langle \mu_1 S_1 || \hat{X}^S || \mu_2 S_2 \rangle
\end{align}
 The full matrix elements are obtained from the reduced matrix elements by
the Wigner-Eckart theorem (analogously to Eq. (\ref{eq:reduced_wf}))
\begin{align}
\mathbf{X}^{Sm}_{\mu_1 S_1 m_1 \mu_2 S_2 m_2} &=  c^{S_2 S S_1}_{m_2 m m_1} \mathbf{X}^S_{\mu_1 S_1 \mu_2 S_2}
 \label{eq:wig1}
\end{align}
The adjoint of a tensor operator is also a tensor operator. Here, we define
the adjoint with a additional sign factor to preserve
the Condon-Shortley phase convention used in the angular momentum
ladder operators. To denote this adjoint with an additional phase, we use the symbol $\ddagger$. For example
\begin{align}
 \mathbf{X}^{S,m \ddagger} = (-1)^{S+m} 
 \mathbf{X}^{S,-m \dag} \label{eq:totran}
\end{align}
Note that  reduced matrix elements of the adjoint of a tensor operator are not the
adjoint of the reduced matrix elements of the operator. The relationship
between the reduced  matrix elements of the tensor operators of spin
$S=0, \frac{1}{2}, 1$ and those of the corresponding adjoint operators, is given in Appendix \ref{sec:transpose}.

As is the case for spin eigenstates, a product tensor operator  with  quantum numbers $S, m$
consists  of a linear combination  of  tensor operators with quantum numbers $S_1,m_1$ and $S_2,m_2$,
coupled through  Clebsch-Gordan coefficients 
\begin{align}
(\hat{X}_1^{S_1}\hat{ X}_2^{S_2})^{S m} &= \sum_{m_1 m_2} c^{S_1 S_2 S}_{m_1 m_2 m}
\hat{X}_1^{S_1 m_1} \hat{X}_2^{S_2 m_2} \label{eq:tensor}
\end{align}
We can obtain the reduced matrix elements of the product operator $(\hat{X}_1^{S_1}\hat{ X}_2^{S_2})^S$
directly from  the reduced matrix elements of the  operators $\hat{X}$ and $\hat{Y}$
using  Wigner 9$j$ coefficients
\begin{align}
& \langle \mu \nu S_{\mu\nu} ||(\hat{X}_1^{S_1}
   \hat{X}_2^{S_2})^{S}||\mu' \nu'
 S_{\mu'\nu'} \rangle \notag  \\
&=
\left[ \begin{array}{ccc} S_{\mu'}& S_{\nu'}& S_{\mu\nu'}\\ S_{1}& S_{2}& S \\ S_\mu&
    S_\nu& S_{\mu\nu}\\
       \end{array}\right] \langle \mu S_\mu||X_1^{S_{1}}||\mu'S_{\mu'}\rangle \langle \nu S_\nu||X_2^{S_{2}}||\nu' S_{\nu'}\rangle  
\label{eq:9j}
\end{align}
Here we define the spin-adapted tensor product $\otimes_S$ as
\begin{align}
\left(\mathbf{X}_1^{S_1} \mathbf{X}_2^{S_2}\right)^S = \mathbf{X}_1^{S_1} \otimes_S \mathbf{X}_2^{S_2} \label{eq:otimes_s}
\end{align}
which is the reduced matrix analogue of Eq. (\ref{eq:tensor}) and the reduced matrix elements of $\left(\mathbf{X}_1^{S_1} \mathbf{X}_2^{S_2}\right)^S$ are calculated as shown in Eq. (\ref{eq:9j}).

We now proceed to discuss how the spin algebra established above can
be applied to the computations of the sweep iteration.

\subsection{Spin-adapted sweep iteration}
\textit{Blocking ---} The two modifications to blocking when implementing spin-adaptation, are (i)
instead of using the operators in Table~\ref{tab:abelian_ops}, we use tensor operators,
defined in Table \ref{tab:ops}, (ii) because we use tensor operators, we manipulate and store only the reduced
matrix elements of the operators. This means that we replace the tensor multiplication $\otimes$, by
the spin-adapted tensor multiplication $\otimes_S$, defined in Eq. (\ref{eq:otimes_s}).

As an example, we consider  the
$A_{ij}^{S}[\mathcal{A}]$ spin tensor operators, whose non-tensor
analogues were considered in Eq. (\ref{eq:aiaj_abblocking}). The matrix of reduced matrix elements corresponding to $A_{ij}^{0}[\mathcal{A}]$ is obtained by
 \begin{align}
 i, j \in \mathcal{L} &\Rightarrow \mathbf{A}_{ij}^0[\mathcal{L}] \otimes_0 \mathbf{1}^0[\bullet_l] \notag \\
 i \in \mathcal{L}, j \in \bullet_l &\Rightarrow \mathbf{a}_i^{1/2}[\mathcal{L}] \otimes_0 \mathbf{a}_j^{1/2}[\bullet_l] \notag \\
 i, j \in \bullet_l &\Rightarrow \mathbf{1}^{0}[\mathcal{L}] \otimes_0 \mathbf{A}_{ij}^0[\bullet_l] 
 \label{eq:aiaj_blocking}
 \end{align}

The partitioning of the superblock Hamiltonian similarly follows
 Eq. (\ref{eq:hblocking}). Here we recall that the Hamiltonian is an $S=0$ operator, i.e. we write $\mathbf{H}^{0}$.
Then
\begin{align} 
&\mathbf{H}^{0}[\mathcal{A}]= \mathbf{H}^{0}[\mathcal{L}] \otimes_0 \mathbf{1}^{0}[\bullet_l] + \mathbf{1}^{0}[\mathcal{L}] \otimes_0 \mathbf{H}^{0}[\bullet_l] \nonumber\\
&+2\sum_{i\in \mathcal{L}} \left(\mathbf{a}_i^{1/2}[\mathcal{L}]\otimes_0 \mathbf{R}_i^{1/2\ddagger}[\bullet_l] + \mathbf{a}_i^{1/2\ddagger}[\mathcal{L}] \otimes_0 \mathbf{R}_i^{1/2}[\bullet_l]\right) \nonumber\\
&+2\sum_{i\in \bullet_l} \left(\mathbf{a}_i^{1/2}[\bullet_l]\otimes_0 \mathbf{R}_i^{1/2\ddagger}[\mathcal{L}] + \mathbf{a}_i^{1/2\ddagger}[\bullet_l] \otimes_0 \mathbf{R}_i^{1/2}[\mathcal{L}]\right) \nonumber\\
&+\sum_{ij\in \bullet_l} \left(-\sqrt{3}\mathbf{B}_{ij}^{1}[\bullet_l]\otimes_0 \mathbf{Q}_{ij}^{1}[\mathcal{L}] + \mathbf{B}_{ij}^{0}[\bullet_l] \otimes_0 \mathbf{Q}_{ij}^{0}[\mathcal{L}]\right) \nonumber\\
&+\frac{\sqrt{3}}{2}\sum_{ij\in \bullet_l} \left(\mathbf{A}_{ij}^{1}[\bullet_l]\otimes_0 \mathbf{P}_{ij}^{1}[\mathcal{L}] + \mathbf{A}_{ij}^{1\ddagger}[\bullet_l] \otimes_0 \mathbf{P}_{ij}^{1\ddagger}[\mathcal{L}]\right) \nonumber\\
&+\frac{1}{2}\sum_{ij\in \bullet_l} \left(\mathbf{A}_{ij}^{0}[\bullet_l]\otimes_0 \mathbf{P}_{ij}^{0}[\mathcal{L}] + \mathbf{A}_{ij}^{0\ddagger}[\bullet_l] \otimes_0 \mathbf{P}_{ij}^{0\ddagger}[\mathcal{L}]\right) \label{eq:hampart}
\end{align}

\begin{table}
 \begin{center}
\caption{Definition of the operators used in the spin-adapted DMRG. Here the indices are spatial
  indices not spin indices.}\label{tab:ops}
  \begin{tabular}{lll}
   \hline \multirow{3}{*}{Label}& \multirow{3}{*}{Operator}& Tensor\\
&& Operator\\
&& Label\\
\hline 
$a_i^{1/2,-1/2}$ &   $a_{i\beta}^\dag$ & \multirow{2}{*}{$a_i^{1/2}$}\\ 
$a_i^{1/2,1/2}$ &   $a_{i\alpha}^\dag$&\\\\ 
$R_k^{1/2,-1/2}$ &   $\frac{1}{\sqrt{2}}\sum_{ijl} \nu_{ijkl}(a_{i\alpha}^\dag
   a_{j\alpha}^\dag a_{l\alpha} + a_{i\alpha}^\dag a_{j\beta}^\dag
   a_{k\beta})$&\multirow{2}{*}{$R_k^{1/2}$}\\ 
$R_k^{1/2,1/2}$ & $\frac{1}{\sqrt{2}}\sum_{ijl}
   \nu_{ijkl}(a_{i\beta}^\dag a_{j\alpha}^\dag a_{k\alpha} +
   a_{i\beta}^\dag a_{j\beta}^\dag a_{k\beta})$&\\\\ 
$A_{ij}^{0,0}$ & $\frac{1}{\sqrt{2}} (a_{i\alpha}^\dag a_{j\beta}^\dag - a_{i\beta}^\dag a_{j\alpha}^\dag)$&$A_{ij}^0$\\\\ 
$A_{ij}^{1,-1}$ & $ a_{i\beta}^\dag a_{j\beta}^\dag$&\multirow{3}{*}{$A_{ij}^{1}$}\\ 
$A_{ij}^{1,0}$ & $\frac{1}{\sqrt{2}} (a_{i\alpha}^\dag a_{j\beta}^\dag + a_{i\beta}^\dag a_{j\alpha}^\dag)$&\\ $A_{ij}^{1,1}$ & $a_{i\alpha}^\dag a_{j\alpha}^\dag$&\\\\ 
$B_{ij}^{0,0}$ & $\frac{1}{\sqrt{2}} (a_{i\alpha}^\dag a_{j\alpha} + a_{i\beta}^\dag a_{j\beta})$&$B_{ij}^0$\\\\
$B_{ij}^{1,-1}$ & $ a_{i\beta}^\dag a_{j\alpha}$&\multirow{3}{*}{$B_{ij}^{1}$}\\ 
$B_{ij}^{1,0}$ & $\frac{1}{\sqrt{2}} (a_{i\alpha}^\dag a_{j\alpha} - a_{i\beta}^\dag a_{j\beta})$&\\ 
$B_{ij}^{1,1}$ & $-a_{i\alpha}^\dag a_{j\beta}$&\\\\
$P_{ij}^{0,0}$ & $\frac{1}{\sqrt{2}} \sum_{kl}-\nu_{ijkl}(-a_{l\alpha} a_{k\beta}+ a_{l\beta} a_{k\alpha})$&$P_{ij}^{0}$\\\\ 
$P_{ij}^{1,-1}$ & $\sum_{kl} \nu_{ijkl}a_{l\alpha} a_{k\alpha}$&\multirow{3}{*}{$P_{ij}^{1}$}\\ 
$P_{ij}^{1,0}$ & $\frac{1}{\sqrt{2}} \sum_{kl}-\nu_{ijkl}(-a_{l\alpha} a_{k\beta} - a_{l\beta} a_{k\alpha})$&\\ $P_{ij}^{1,1}$ & $\sum_{kl}\nu_{ijkl}a_{l\beta} a_{k\beta} $&\\\\ 
$Q_{ij}^{0,0}$ & $\frac{1}{\sqrt{2}} \sum_{kl}(-\nu_{iklj}+2\nu_{ikjl})(a_{k\alpha}^\dag a_{l\alpha} + a_{k\beta}^\dag a_{l\beta})$&$Q_{ij}^{0}$\\\\ 
$Q_{ij}^{1,-1}$ & $ \sum_{kl}-\nu_{iklj}a_{k\beta}^\dag a_{l\alpha} $&\multirow{3}{*}{$Q_{ij}^{1}$}\\ 
$Q_{ij}^{1,0}$ & $\frac{1}{\sqrt{2}} \sum_{kl}-\nu_{iklj}(a_{k\alpha}^\dag a_{l\alpha} - a_{k\beta}^\dag a_{l\beta})$&\\ 
$Q_{ij}^{1,1}$ & $ \sum_{kl}\nu_{iklj}a_{k\alpha}^\dag a_{l\beta} $&\\\\ 
\hline
  \end{tabular}
 \end{center}
\end{table}

\textit{Wavefunction solution} --- In the wavefunction solution step, the spin-adapted
Hamiltonian wavefunction product
 can be performed
entirely in terms of the reduced operator matrix elements and reduced wavefunction coefficients.
As in non-spin adapted DMRG algorithm, the full Hamiltonian matrix is never generated and the product is carried out for each term in the sum in the Hamiltonian in Eq. (\ref{eq:hampart}). For example, Eq. (\ref{eq:hc_intermediates}) becomes 
\begin{align}
&\mathbf{C}_{a' S_a' b'S_b'} = \nonumber\\
&\sum_{S_aS_b}\left[ \begin{array}{ccc} S_b& S_a& S\\ S_{J}& S_{I}&
    0\\ S_b'& S_a'& S'\\
       \end{array}\right]\langle S_b'||\mathbf{O}^{S_J}_{J}[\mathcal{B}]||S_b\rangle
\langle S_a'||\mathbf{O}^{S_I}_{I}[\mathcal{A}]||S_a\rangle \mathbf{C}_{a S_a bS_b} \label{eq:hv}
\end{align}
Note, however, because of the appearance of the 9$j$ coefficients,
the operator product does not separate into two decoupled multiplets, as in the non-spin adapted case shown in Eq. (\ref{eq:hc_intermediates}). This leads to some overhead in the spin-adapted algorithm relative to the non-spin-adapted case, depending on the number of $9j$ coefficients that need to be considered.


\textit{Renormalisation and decimation ---}
In the spin-adapted renormalisation and decimation step we do not seek a simple optimal truncation
of the states of $\mathcal{A}$, but rather an optimal truncation
to a set of states consistent with spin symmetry, i.e. to a set of pure spin states.
These cannot be obtained as eigenvectors of the reduced density matrix of $\mathcal{A}$, 
because it does not commute
with the  spin operator $\hat{S}^2$ of block $\mathcal{A}$. 
As shown in McCulloch \textit{et al.}~\cite{mcculloch1}, the density matrix to use
in this case is the quasi-density matrix, which is obtained from the usual
 density matrix by setting off-diagonal
blocks, that couple states of different spins,  to
zero.
All operations of the renormalisation and decimation step
can be carried out in the multiplet representation, working in terms of  reduced wavefunction 
coefficients and  reduced  matrix elements. The reduced matrix elements of
the quasi-density matrix are obtained from the reduced wavefunction coefficients.
\begin{align}
\mathbf{\Gamma}_{aS_a, a'S_a} = \sum_{b S_b} \mathbf{C}_{a S_a bS_b} \mathbf{C}^*_{a' S_a bS_b} 
\end{align}
The eigenvectors of the quasi-density matrix yield the transformation matrices in reduced form, via its eigenvectors
\begin{align}
\hat{\Gamma} ||l_{S}\rangle = \sigma_{l,S} ||l_{S}\rangle
\end{align}
After obtaining the new renormalized basis, the operators in multiplet representation are transformed using the analogous formula to Eq. (\ref{eq:optran}).

Note that when  retaining $M$ eigenvectors of the quasi-density matrix in the 
multiplet representation, we are retaining $M$ sets of spin-multiplets.
This corresponds to a much larger set of underlying states, which
is of course, the advantage of working in a spin-adapted formulation. However we will still use the terminology ``$M$ states'' to refer to the renormalized basis in the spin-adapted algorithm.

As described in the non-spin adapted case, the convergence of the Davidson iteration
is greatly improved if we use a suitable guess obtained by transforming
 the wavefunction from the previous sweep iteration. The transformation of the wavefunction in the case of the spin-adapted algorithm is very similar to the case of the non-spin-adapted algorithm with the exception that a spin-rescaling step must be performed, involving the Racah coefficients. Eq. (\ref{eq:firststep}) is analogous to Eq. (\ref{eq:wave_transform}), but in the last step we explicitly specify the spin quantum number of each multiplet state because these are required in the Racah coefficients $W$. In Eq. (\ref{eq:wave_transform2}) instead of matrix coefficients we use the bra-ket notation, to explicitly show the couplings of the spins, so for example $\mathbf{G}_{l_{p}, n_{p+1} n_{p+2}r_{p+3}}$ is the same as $\langle l_{p;S_1} ||\langle n_{p+1;S_2} \left(n_{p+2;S_3}r_{p+3;S_4}(S_5)\right)(S_{25})||\Psi(S)\rangle$, where states $n_{p+2;S_3}$ and $r_{p+3;S_4}$ couple to form a state with spin $S_5$, which in turn couples to state $n_{p+1;S_2}$ to form a state with spin $S_{25}$.

\begin{widetext}
\begin{align}
\mathbf{G}_{l_{p}, n_{p+1} n_{p+2}r_{p+3}} =&\sum_{\substack{l_{p-1}n_p \\ r_{p+1}}}  \tilde{\mathbf{L}}_{l_p,l_{p-1} n_{p}} \mathbf{R}_{n_{p+2} r_{p+3},r_{p+2}} \mathbf{C}_{l_{p-1} n_{p}, n_{p+1}r_{p+2}}  \label{eq:firststep}
\end{align}
\begin{align}
\langle l_{p;S_1}  n_{p+1;S_2}(S_{12})&||\langle n_{p+2;S_3}r_{p+3;S_4}(S_5)||\Psi(S)\rangle =\nonumber\\
&\sum_{\substack{l_{p}  n_{p+1} \\  r_{p+2}}} W(S_1S_2SS_5; S_{12} S_{25})\left[(2S_{12}+1)(2S_{23}+1)\right]^{1/2}\times\nonumber\\
&  \langle l_{p;S_1} ||\langle n_{p+1;S_2} \left(n_{p+2;S_3}r_{p+3;S_4}(S_5)\right)(S_{25})||\Psi(S)\rangle
\label{eq:wave_transform2}
\end{align}
\end{widetext}


\section{Computational considerations}

\label{sec:computational}

The  computational implementation of the spin-adapted DMRG algorithm is 
 similar to the non-spin-adapted DMRG. Here we focus  on computational differences
between the two. 
\begin{itemize}
\item The total number of operators stored in the spin-adapted DMRG is
approximately half that in the non-spin-adapted DMRG. The most numerous kinds of operators in 
the DMRG algorithm are those with
 two orbital indices $i$ and $j$, namely $\hat{A}_{ij}, \hat{B}_{ij}, \hat{P}_{ij}, \hat{Q}_{ij}$.
 In the non-spin-adapted case 
there are four different $\hat{A}_{ij}$ operators for every spatial pair $ij$, i.e.
 $\hat{A}_{i\alpha j\alpha}$, $\hat{A}_{i\beta j\alpha}$, $\hat{A}_{i\alpha j\beta}$, and $\hat{A}_{i\beta j\beta}$. 
In the spin-adapted case, there are only two tensor operators: $\hat{A}_{ij}^{0}$ and $\hat{A}_{ij}^{1}$. $\hat{A}_{ij}^1$ contains
three $m$ components, but the Wigner-Eckart theorem (Eq. (\ref{eq:wig1})) means we store
only a \textit{single} matrix of reduced matrix elements.
\item 
The storage dependence of the spin-adapted algorithm is $O(M^2)$ which is the same scaling as in the non-spin-adapted algorithm. 
However, the storage prefactor in the spin-adapted case is larger. 
This arises from the non-Abelian nature of the spin symmetry.
For example, if we consider  an operator such as  $\hat{B}_{ij}^{1}$,  the following reduced matrix elements
  are non-zero: $\langle\mu_1 
  S||\hat{B}_{ij}^1||\mu_2 S-1\rangle$, $\langle\mu_1
  S||\hat{B}_{ij}^1||\mu_2 S\rangle$ and $\langle\mu_1
  S||\hat{B}_{ij}^1||\mu_2 S+1\rangle$ i.e. several different couplings between bra and ket are allowed. 
When Abelian symmetries are used, 
  $\hat{B}_{i\alpha j\beta}$ has  non-zero matrix
  elements only between states of a single type $\langle \mu_1  m|$ and
  $|\mu_2  m\rangle$. 

\item The main cost of the algorithm comes from the  Hamiltonian
wavefunction multiplication in the wavefunction solution step, and  the operator transformation, in
the renormalisation and decimation step. In the spin-adapted
case, the  cost of the Hamiltonian
wavefunction multiplication is $O(k^2M^3)$ per sweep step, similar to the non-spin-adapted algorithm. In the spin-adapted algorithm the presence of the $9j$ coupling coefficients prevents the Hamiltonian wavefunction multiplication from factoring into two stages as in Eq. (\ref{eq:hc_intermediates}). The prefactor of this step thus depends on the number of $9j$ couplings that must be accounted for. For singlet states, the spin-adapted computational prefactor is similar to that of the non-spin-adapted case but for higher spin states, it can be larger. The operator transformation in the spin-adapted algorithm is very similar to the non-spin-adapted case (and scales as $O(k^2M^3)$ per sweep step) except for the fact that some of the operators are more dense as described in the previous paragraph. 

\item For large scale calculations an efficient parallelization of the
  code is required. We have carried this out in the exact same way as in
  the non-spin-adapted DMRG algorithm described by Chan\cite{Chan2004}.

\end{itemize}

\subsection{Singlet Embedding}\label{sec:singlet_embedding}

When using the spin-adapted DMRG algorithm to study higher spin states than the singlet, some disadvantages appear.
Firstly, the reduced coefficient matrix $\mathbf{C}_{aS_a bS_b}$ becomes more dense. In the case of
the singlet, only quantum states of equal spins on blocks $\mathcal{A}$ and $\mathcal{B}$ can couple, while for
say, a triplet state, additional couplings ($S_b=S_a\pm 1$) are possible. A second disadvantage (related to the first)
is that for non-singlet states, the eigenvalues of the quasi-density matrix of block $\mathcal{A}$ and of block $\mathcal{B}$ are not equivalent.
A simple example illustrates this. Consider
a reduced wavefunction written as
\begin{align}
||\Psi_{S=1}\rangle= \frac{1}{\sqrt{2}}||aS_a=1\rangle (||bS_b=0\rangle + ||bS_b=2\rangle)
\end{align}
The quasi-density matrix of block $\mathcal{A}$ has one non-zero eigenvalue, while that of block $\mathcal{B}$ has two
non-zero eigenvalues. This non-equivalence means that  discarded weights obtained during the forward and
backward sweeps of a calculation (which respectively
arise from quasi-density matrices of blocks $\mathcal{A}$ and $\mathcal{B}$) are different, and this makes
DMRG energy extrapolation using discarded weights ambiguous.

To overcome these disadvantages, it is clearly best to use the spin-adapted algorithm only to target singlet states. How then
do we study systems in a higher spin state? One way is to use a technique which we call singlet embedding, originally introduced
by Nishino \textit{et al.}\cite{tatsuaki}. Here we note that we can always add a set of auxiliary non-interacting orbitals to the end of the lattice which
couple to the physical orbitals to overall yield a singlet state. In general, the wavefunction $||\tilde{\Psi}\rangle$ of
the combined physical and auxiliary orbitals is of the form
\begin{align}
||\tilde{\Psi}_{S=0}\rangle = ||\Psi_S\rangle ||\Phi_S\rangle
\end{align}
where $||\Phi_S\rangle$ is the state of the auxiliary non-interacting orbitals. Because the auxiliary orbitals
do not energetically couple to the physical system, and have themselves no energy, they
 do not affect the energy of the physical system. We have implemented the singlet embedding technique as an option
in our calculations, as described below.

\section{Applications}\label{sec:applications}


In this section we describe application of spin-adapted DMRG algorithm to study two small transition metal complexes, \fes~\cite{fe2s21,fe2s22} and \Cr~\cite{celani,Andersson,Mitrushenkov} which have been of interest in quantum chemistry. In the first calculation, we target the  spin ladder of the \fes molecule. 
Here we use  a small active space of 12 electrons in 12 orbitals ($12e,12o$)
and a minimal basis\cite{sto3g}, to demonstrate the ability of the spin-adapted DMRG algorithm to
 target very closely spaced states of different spatial and spin symmetries. In our second calculation, we
study the singlet and triplet spin states of the \Cr molecule. This is a benchmark calculation 
using a large active space  ($24e,30o$) but a small (single-valence) basis set, that follows
closely the earlier work of Kurashige and Yanai on the same system. This calculation is primarily intended
to examine the relative efficiencies of the spin-adapted and non-spin-adapted algorithms.

\subsection{Fe$_2$S$_2$}
We first carried out spin-adapted DMRG calculations on the \fes molecule. The geometry, which exhibits
 $D_{2h}$ point group symmetry, is  given in
 Table~\ref{tab:stut1}. We used a minimal STO-3G\cite{sto3g} basis. The active space was identified by carrying out 
a high-spin UB3LYP/STO-3G\cite{b3,lyp,sto3g} calculation with multiplicity 9 (eight unpaired electrons), and then selecting 12 unrestricted natural orbitals with 
occupation numbers between 1.99 and 0.01  to make up the (12$e$, 12$o$) active space. 
The order of the orbitals in the DMRG calculation was by occupation number. We then carried out calculations on 40 states (multiplicities 1, 3, 5, 7, 9, for each of the 8 irreps of $D_{2h}$). With M=200 the DMRG energies (in $E_h$) were already converged to 5 decimal places as compared to the ORCA\cite{orca} complete active space configuration interaction (CASCI) results.

The corresponding energies are given in Table \ref{tab:ene}. As  can be seen, many of the states
are nearly degenerate (to within $<$10$\mu$H) and thus would be extremely hard to resolve without
a spin-adapted algorithm.



\begin{table}
 \caption{Cartesian coordinates of Fe$_2$S$_2$ with $D_{2h}$ symmetry. The molecule
lies in the $yz$ plane.}
 \begin{tabular}{ccc}
  \hline
\hline
  \multirow{2}{*}{Atom} &y  &z  \\
  &\multicolumn{2}{c}{\AA}\\
  \hline
     S  &           0.00 &  -4.29\\
     S  &           0.00 &   4.29\\
     Fe &          -2.10 &   0.00\\
     Fe &           2.10 &   0.00\\
     \hline
\hline
\end{tabular}
\label{tab:stut1}
\end{table}

\begin{widetext}
\begin{center}
\begin{table}[hpt]
\caption{Energies ($E+3283.0$) in $E_h$ of \fes in various spin and symmetry states calculated using the spin-adapted DMRG algorithm. The results agree with ORCA CASCI energies to all the decimal places shown. Note the very close spacing of the states, which would be very hard to resolve without a spin-adapted algorithm.}
\begin{tabular}{cccccc}
\hline
\hline
\multirow{2}{*}{Irrep}&\multicolumn{5}{c}{Multiplicity}\\
&1&3&5&7&9\\
\hline
$A_g$&-0.75990&	-0.75990&	-0.75993&	-0.75993&	-0.75996\\
$B_{1g}$&-0.75992&	-0.75992&	-0.75991&	-0.75993&	-0.75996\\
$B_{2g}$&-0.77343&	-0.78351&	-0.78343&	-0.72207&	-0.78312\\
$B_{3g}$&-0.77345&	-0.78007&	-0.77991&	-0.78662&	-0.78648\\
$A_u$&-0.76308&	-0.77344&	-0.78678&	-0.78669&	-0.78656\\
$A_{1u}$&-0.77686&	-0.77672&	-0.78333&	-0.72207&	-0.78301\\
$A_{2u}$&-0.68761&	-0.75991&	-0.75990&	-0.75992&	-0.75995\\
$A_{3u}$&-0.69535&	-0.75991&	-0.75993&	-0.69537&	-0.69614\\
\hline
\hline
\end{tabular}
\label{tab:ene}
\end{table}
\end{center}
\end{widetext}

\subsection{\Cr}

Recently  Kurashige and Yanai\cite{kurashige} carried out
large-scale DMRG calculations on the singlet ground state of Cr$_2$ using an active space
of $(24e, 30o)$. These were benchmark rather than realistic calculations because they used a small single valence (SV) basis
set which did not include dynamical correlation (see however Ref. \cite{kurashige_new} for a more
detailed DMRG with perturbation theory study  of the chromium dimer with the inclusion of dynamical correlation).
Here, we use the same Cr$_2$ benchmark example  as Kurashige and Yanai
with exactly the same geometry (bond length 1.5~\AA), molecular  
orbitals and ordering as in their original paper. Our purpose 
 will be to examine the accuracy and speed of the spin-adapted DMRG algorithm
 as compared to the non-spin-adapted algorithm.  
We target the singlet ($^1A_g$ in $D_{2h}$ symmetry) and triplet ($^3B_{1g}$ in $D_{2h}$ symmetry) states of the molecule in our calculations.

\subsubsection{Accuracy}

\begin{center}
\begin{table}
\caption{Energy in $E_h$ and discarded weights of a spin-adapted DMRG calculation on the singlet state of the \Cr molecule. Note that our M=5000 spin-adapted energy is already better than the M=10000 non-spin-adapted energy reported by Kurashige and Yanai~\cite{kurashige}.}
\begin{tabular}{ccc}
\hline
\hline
M&Energy($E_h$)&Discarded weight\\
\hline
1000	&-2086.41831&	$2.032\times10^{-5}$\\
2000	&-2086.41979&	$1.006\times10^{-5}$\\
5000	&-2086.42061&	$3.075\times10^{-6}$\\
8000	&-2086.42078&	$1.608\times10^{-6}$\\
10000&	-2086.42082&	$9.630\times10^{-7}$\\
$\infty$&-2086.42100&\\
\hline
\hline
\end {tabular}
\label{tab:cr2a}
\end{table}
\end{center}

The total DMRG energy of the singlet state as a function of the number of retained states (M),
as well as the   discarded weight in the quasi-density matrix (the largest discarded weight during the DMRG sweep), is shown in Table~\ref{tab:cr2a}.
Kurashige and Yanai's converged DMRG energy with 10000 non-spin-adapted states was $-2086.42053$ $E_h$ which is slightly \emph{above}
our spin-adapted M=5000 energy of $-2086.42061$ $E_h$. We see that the spin-adapted DMRG algorithm requires 
roughly only half the number of states as the non-spin-adapted DMRG, to achieve a similar accuracy in the energy. The greater
accuracy of the spin-adapted algorithm allows us to perform a more accurate extrapolation of the DMRG energy
to $M=\infty$ than in~\cite{kurashige} and our final $M=10000$ spin-adapted DMRG energy is within 0.2 m$E_h$ of the extrapolated $M=\infty$ result.

\begin{figure}
\begin{center}
\resizebox{80mm}{!}{\includegraphics{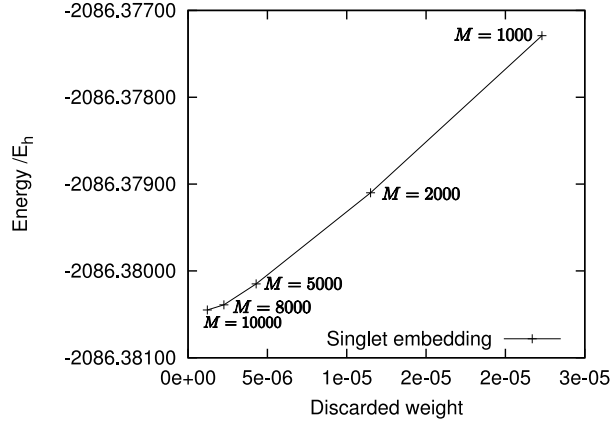}}
\end{center}
\caption{DMRG energy in $E_h$ of the \Cr triplet state on the y-axis versus discarded weight on the x-axis using the singlet embedding approach.}
\label{fig:extrap}
\end{figure}

The total DMRG energy and the discarded weights of the triplet state using the 
spin-adapted (with and without spin embedding) and non-spin-adapted algorithms are shown in 
Table~\ref{tab:cr22}. Similarly to the singlet case, we find that the spin-adapted algorithm
requires roughly half the number of renormalised states as the non-spin-adapted algorithm
to achieve the same accuracy. Singlet embedding (section \ref{sec:singlet_embedding}), although
formally increasing the number of orbitals in the problem, leads to no loss of accuracy
as compared to the spin-adapted calculation on the triplet state, and indeed leads to a slight increase in accuracy.
As observed in section \ref{sec:singlet_embedding}, in the spin-adapted calculation on 
the triplet state, the discarded weights obtained during the forward and backward sweeps
are vastly different. This discrepancy vanishes when the triplet state energies
are obtained via embedding in a singlet state. The singlet embedding allows us to perform energy extrapolation
with respect to the discarded weights, as shown in Fig. \ref{fig:extrap}. We find that the $M=10000$
spin-adapted calculation is within 0.3 mH of the extrapolated exact DMRG result.


\begin{widetext}
\begin{center}
\squeezetable
\begin{table}[thp]
\caption{Energies in $E_h$ and discarded weights of a spin-adapted DMRG calculation on the triplet state of the \Cr molecule. Columns 2 through 5 give data for the forward and backward sweeps of spin-adapted calculations, columns 6 and 7 give our results when we use the singlet embedding technique and finally the last two columns give our results of non-spin-adapted calculations.} 
{
\begin{tabular}{ccccccccc}
\hline
\hline
M&\multicolumn{6}{c}{Spin-adapted DMRG}&\multicolumn{2}{c}{Non-spin-adapted}\\
&\multicolumn{2}{c}{Forward sweep}&\multicolumn{2}{c}{Backward sweep}&\multicolumn{2}{c}{Singlet embedding}&&\\
&Energy($E_h$)&Discarded weight&Energy($E_h$)&Discarded weight&Energy($E_h$)&Discarded weight&Energy($E_h$)&Discarded weight\\
\hline
1000	&-2086.37682&	$1.45\times 10^{-4}$&-2086.37682&$1.77\times 10^{-5}$&-2086.37729&$2.23\times 10^{-5}$&-2086.37418&$5.89\times 10^{-5}$\\
2000	&-2086.37888&	$6.67\times 10^{-5}$&-2086.37888&$1.10\times 10^{-5}$&-2086.37910&$1.15\times 10^{-5}$&-2086.37736&$2.75\times 10^{-5}$\\
5000	&-2086.38011&	$2.46\times 10^{-5}$&-2086.38009&$4.69\times 10^{-6}$&-2086.38015&$4.29\times 10^{-6}$&-2086.37949&$1.10\times 10^{-5}$\\
8000	&-2086.38036&	$1.16\times 10^{-5}$&-2086.38036&$2.78\times 10^{-6}$&-2086.38039&$2.26\times 10^{-6}$&-2086.38000&$5.86\times 10^{-6}$\\
10000   &-2086.38043&	$8.56\times 10^{-6}$&-2086.38043&$1.94\times 10^{-6}$&-2086.38045&$1.21\times 10^{-6}$&-2086.38016&$3.52\times 10^{-6}$\\
$\infty$&&&&&-2086.38074&&-2086.38059&\\
\hline
\hline
\end {tabular}
}\label{tab:cr22}
\end{table}
\end{center}
\end{widetext}


\subsubsection{Efficiency}

As explained in Section~\ref{sec:computational} the most expensive step in the DMRG algorithm
is formation of the Hamiltonian wavefunction product, whose computational cost scales as
 $O(M^3)$, where $M$ is the number of retained states.
From the above results, we observe that the spin-adapted algorithm requires roughly half the
number of renormalised states as the non-spin-adapted algorithm to achieve the same accuracy.
This suggests that if the cost of a single Davidson iteration (for a given number of states)
is comparable between the spin-adapted and non-spin-adapted algorithms, then, 
to achieve a given accuracy in the DMRG energy, the spin-adapted algorithm should offer an 8-fold
gain in computational speed.

To compare the performance of the spin-adapted and non-spin-adapted DMRG algorithms we show the wall times per Davidson iteration of the two algorithms  in Table~\ref{tab:cpu}. For the singlet case, we notice that for example the $M=5000$ timings are 
comparable for both the spin-adapted and non-spin-adapted calculations. However, moving to $M=10000$, the
computational cost increases by a factor of 4 rather than 8 i.e. more like $O(M^2)$ rather than $O(M^3)$. This means that the spin-adapted algorithm
yields (for a given accuracy) only a 4-fold gain in computational efficiency over the non-spin-adapted algorithm.
The quadratic scaling is a result of the high Abelian spatial symmetry ($D_{2h}$) present in the molecule,
which means that each of the non-zero blocks of the operators are so small that the corresponding BLAS 
matrix multiplication operations are dominated by quadratic as opposed to cubic complexity terms.
We expect, however, the computational scaling would approach $O(M^3)$ as $M$ is increased further, or if the calculations were performed without the use of point group symmetry, as may be the case in other more complex molecules, in which case the spin-adapted algorithm should offer even larger computational gains.

In the triplet state, as expected from the analysis in section \ref{sec:computational}, for
any given $M$, the cost of the Davidson iteration is much higher for the spin-adapted algorithm
than for the non-spin-adapted algorithm. However, with singlet embedding, the spin-adapted computational
times are now similar to those of the non-spin-adapted case. Thus, with singlet embedding, the spin-adapted
algorithm also provides a 4-fold efficiency gain for the triplet state, which we expect to rise
either as $M$ is increased, or if we consider more complex molecules without high Abelian spatial symmetry.


\begin{center}
\begin{table}
\caption{Wall clock times for a single Davidson iteration performed with spin-adapted and non-spin-adapted DMRG algorithms on 2 Intel Xeon E5420 processors with 4 cores each. Spin$_{se}$ is the singlet embedding technique where we add a set of non-interacting orbitals and then target the $S=0$ state of the combined system (see text for more details). Ratio denotes the ratio of the best spin-adapted to non-spin-adapted timings. Note that the singlet embedding technique greatly reduces the cost of the spin-adapted DMRG calculation for the triplet state.} 
\begin{tabular}{cccc|cccc}
\hline
&\multicolumn{2}{c}{S=0}&&\multicolumn{3}{|c}{S=1}&\\
M&\multicolumn{2}{c}{Timings /s}&&\multicolumn{3}{|c}{Timings /s}&\\
&Spin&non-Spin&Ratio&Spin&Spin$_{se}$&non-Spin&Ratio\\
\hline
2000	&59&55&1.07&111&41&48&0.85\\
5000	&329&292&1.13&707&248&267&0.93\\
8000	&1003&794&1.26&2622&792&746&1.06\\
10000   &1752&1363&1.29&4628&1782&1295&1.38\\
\hline
\end {tabular}
\label{tab:cpu}
\end{table}
\end{center}

\section{Conclusions}
In this work we implemented a spin-adapted density matrix renormalization group algorithm that
extends the work of McCulloch and Gulacsi to quantum chemical Hamiltonians. The spin-adapted algorithm opens up the
individual targetting
of closely spaced spin states. Furthermore, when used in conjunction with
the singlet embedding technique of Nishino, we find that significant computational gains are possible. In the
systems studied here,
the number of  spin-adapted renormalised states required for a given accuracy is roughly only half
that of the non-spin-adapted renormalised states. This yields a theoretical computational speedup of a factor of 8,
although we observe speedups closer to 4 due to the high point group symmetry of the systems we have studied.
The ability to target individual spin states,
as well as the improved computational efficiency of the spin-adapted algorithm, will be particularly advantageous
when studying larger transition metal complexes such as those which involve multiple metal centres. Such studies
are currently in progress in our group.

\section{Acknowledgements}
This work was supported by an NSF CHE CAREER grant, NSF-CHE-0645380.

\appendix
\section{Blocking}\label{sec:appendix_blocking}
In this section we give the formulae for formation of operators $R$, $P$ and $Q$ in the blocking step of non-spin-adapted and spin-adapted DMRG.
\subsection{Non-spin-adapted DMRG}
\begin{align}
\mathbf{R}_i[\mathcal{A}] =& \mathbf{R}_i[\mathcal{L}] \otimes \mathbf{1}[\bullet_l] + \mathbf{R}_i[\bullet_l] \otimes \mathbf{1}[\mathcal{L}]\nonumber\\
&+ \sum_{j\in \bullet_l} 2\mathbf{P}_{ij}[\mathcal{L}]\otimes \mathbf{a}_j^{\dag}[\bullet_l] + \mathbf{Q}_{ij}[\mathcal{L}] \otimes \mathbf{a}_i[\bullet_l] \nonumber\\
&+ \sum_{j\in \mathcal{L}} 2\mathbf{P}_{ij}[\bullet_l]\otimes \mathbf{a}_j^{\dag}[\mathcal{L}] + \mathbf{Q}_{ij}[\bullet_l] \otimes \mathbf{a}_i[\mathcal{L}] 
\label{eq:rblocking}
\end{align}

\begin{align}
\mathbf{Q}_{ij}[\mathcal{A}] =& \mathbf{Q}_{ij}[\mathcal{L}] \otimes \mathbf{1}[\bullet_l] + \mathbf{Q}_{ij}[\bullet_l] \otimes \mathbf{1}[\mathcal{L}]\nonumber\\
&+ 2\sum_{\substack{ k\in \bullet_l \\l \in\mathcal{L}} }\left( (v_{ikjl}-v_{iklj})\mathbf{a}_{k}^{\dag}[\bullet_l] \otimes \mathbf{a}_l[\mathcal{L}]\right. \nonumber\\
&\left.+ (v_{iljk}-v_{ilkj})\mathbf{a}_{k}[\bullet_l]\otimes \mathbf{a}_l^{\dag}[\mathcal{L}]\right)
\label{eq:qblocking}
\end{align}

\begin{align}
\mathbf{P}_{ij}[\mathcal{A}] =& \mathbf{P}_{ij}[\mathcal{L}] \otimes \mathbf{1}[\bullet_l] + \mathbf{P}_{ij}[\bullet_l] \otimes \mathbf{1}[\mathcal{L}]\nonumber\\
&+ \sum_{\substack{ k\in \bullet_l\\l \in\mathcal{L}} }\left(v_{ijlk}\mathbf{a}_{k}[\bullet_l]\otimes  \mathbf{a}_l[\mathcal{L}]+ v_{ijkl}\mathbf{a}_{k}[\bullet_l]\otimes \mathbf{a}_l[\mathcal{L}]\right)
\label{eq:pblocking}
\end{align}

\subsection{Spin-adapted DMRG}
\begin{align}
&\mathbf{R}_i^{1/2}[\mathcal{A}] = \mathbf{R}_i^{1/2}[\mathcal{L}] \otimes_{1/2} \mathbf{1}^{0}[\bullet_l] + \mathbf{R}_i^{1/2}[\bullet_l] \otimes_{1/2} \mathbf{1}^{0}[\mathcal{L}]\nonumber\\
&+ \sum_{j\in \bullet_l} \frac{\sqrt{3}}{2}\mathbf{P}_{ji}^{1}[\mathcal{L}]\otimes_{1/2} \mathbf{a}_j^{1/2}[\bullet_l] + \frac{1}{2}\mathbf{P}_{ji}^{0}[\mathcal{L}] \otimes_{1/2}\mathbf{a}_i^{1/2}[\bullet_l] \nonumber\\
&+ \sum_{j\in \bullet_l} \frac{\sqrt{3}}{2}\mathbf{Q}_{ij}^{1\ddagger}[\mathcal{L}]\otimes_{1/2} \mathbf{a}_j^{1/2\ddagger}[\bullet_l] - \frac{1}{2}\mathbf{Q}_{ij}^{0\ddagger}[\mathcal{L}] \otimes_{1/2}\mathbf{a}_i^{1/2\ddagger}[\bullet_l] \nonumber\\
&+ \sum_{j\in \mathcal{L}} \frac{\sqrt{3}}{2}\mathbf{P}_{ji}^{1}[\bullet_l]\otimes_{1/2} \mathbf{a}_j^{1/2}[\mathcal{L}] + \frac{1}{2}\mathbf{P}_{ji}^{0}[\bullet_l] \otimes_{1/2}\mathbf{a}_i^{1/2}[\mathcal{L}] \nonumber\\
&+ \sum_{j\in \mathcal{L}} \frac{\sqrt{3}}{2}\mathbf{Q}_{ij}^{1\ddagger}[\bullet_l]\otimes_{1/2} \mathbf{a}_j^{1/2\ddagger}[\mathcal{L}] - \frac{1}{2}\mathbf{Q}_{ij}^{0\ddagger}[\bullet_l] \otimes_{1/2}\mathbf{a}_i^{1/2\ddagger}[\mathcal{L}] \nonumber\\
\label{eq:rblocking_sp}
\end{align}

\begin{align}
\mathbf{Q}_{ij}^{1}[\mathcal{A}] =& \mathbf{Q}_{ij}^{1}[\mathcal{L}] \otimes_1 \mathbf{1}^{0}[\bullet_l] + \mathbf{Q}_{ij}^{1}[\bullet_l] \otimes_1 \mathbf{1}^{0}[\mathcal{L}]\nonumber\\
 -\sum_{\substack{ l\in \bullet_l\\ k\in\mathcal{L}} }&\left( v_{kijl} \mathbf{a}_{l}^{1/2;\ddagger}[\bullet_l] \otimes_{1}\mathbf{a}_k^{1/2}[\mathcal{L}] \right.\nonumber\\
&\left.+ v_{lijk} \mathbf{a}_{l}^{1/2}[\bullet_l] \otimes_1\mathbf{a}_l^{1/2\ddagger}[\mathcal{L}]\right)
\label{eq:q1blocking_sp}
\end{align}

\begin{align}
\mathbf{Q}_{ij}^{0}[\mathcal{A}] =& \mathbf{Q}_{ij}^{0}[\mathcal{L}] \otimes_0 \mathbf{1}^{0}[\bullet_l] + \mathbf{Q}_{ij}^{0}[\bullet_l] \otimes_0 \mathbf{1}^{0}[\mathcal{L}]\nonumber\\
 -\sum_{\substack{ l\in \bullet_l\\ k\in\mathcal{L}} }&\left( (2v_{ikjl}-v_{kijl}) \mathbf{a}_{l}^{1/2;\ddagger}[\bullet_l] \otimes_0\mathbf{a}_k^{1/2}[\mathcal{L}] \right.\nonumber\\
&\left.+ (2v_{iljk}-v_{lijk}) \mathbf{a}_{l}^{1/2}[\bullet_l] \otimes_0\mathbf{a}_l^{1/2\ddagger}[\mathcal{L}]\right)
\label{eq:q0blocking_sp}
\end{align}

\begin{align}
\mathbf{P}_{ij}^{1}[\mathcal{A}] =& \mathbf{P}_{ij}^{1}[\mathcal{L}] \otimes_1 \mathbf{1}^{0}[\bullet_l] + \mathbf{P}_{ij}^{1}[\bullet_l] \otimes_1 \mathbf{1}^{0}[\mathcal{L}]\nonumber\\
&- \sum_{\substack{ k\in\mathcal{L} \\l \bullet_l\in} } v_{ijlk}\mathbf{a}_{k}^{1/2\ddagger}[\bullet_l] \otimes_1 \mathbf{a}_l^{1/2\ddagger}[\mathcal{L}]
\label{eq:p1blocking_sp}
\end{align}

\begin{align}
\mathbf{P}_{ij}^{0}[\mathcal{A}] =& \mathbf{P}_{ij}^{0}[\mathcal{L}] \otimes_1 \mathbf{1}^{0}[\bullet_l] + \mathbf{P}_{ij}^{0}[\bullet_l] \otimes_1 \mathbf{1}^{0}[\mathcal{L}]\nonumber\\
&+ \sum_{\substack{ k\in\mathcal{L} \\l \bullet_l\in} } (v_{ijlk}-v_{ijkl})\mathbf{a}_{k}^{1/2\ddagger}[\bullet_l] \otimes_0 \mathbf{a}_l^{1/2\ddagger}[\mathcal{L}]
\label{eq:p0blocking_sp}
\end{align}

\section{Matrix Product State formulation}~\label{sec:mps}
\begin{center}
 \begin{figure}
\resizebox{80mm}{!}{\includegraphics{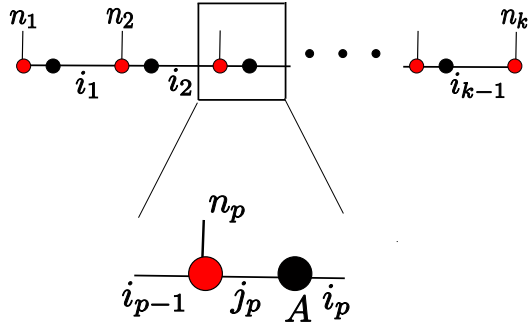}}
\caption{The figure shows the graphical representation of the MPS wavefunction that can be obtained using the results of the spin-adpated DMRG calculation. The red dots represent a matrix of Clebsch-Gordan coefficients ($\mathbf{U}^n$) and the black dots are the rotation matrices obtained from the renormalization step in DMRG ($\underline{\mathbf{L}}, \underline{\mathbf{R}}$ see text for details).}
\label{fig:mps}
\end{figure}
\end{center}

The wavefunction emerging from the usual non-spin-adapted DMRG has a matrix product state (MPS) structure
 as described in many references~\cite{Ostlund1995,Schollwock2011,chan2011}. In the canonical form associated
with a given block configuration, the MPS wavefunction is written as (using the one-dot
formulation of the DMRG for simplicity ~\cite{Ostlund1995,Schollwock2011})
\begin{align}
|\Psi\rangle = \sum_{\{\mathbf{n}\}} \mathbf{L}^{n_1} \mathbf{L}^{n_2} \ldots \mathbf{C}^{n_p} \mathbf{R}^{n_{p+1}} \ldots \mathbf{R}^{n_k} |\mathbf{n}\rangle
\end{align}
where $|\mathbf{n}\rangle$ denotes a Slater determinant in occupation number representation, $\mathbf{L}^{n}$
is a left transformation matrix as defined in Eq. (\ref{eq:lcoeffs}), obtained during the forwards DMRG sweep,  $\mathbf{R}^n$
is a right transformation matrix, obtained during the backwards sweep, and $\mathbf{C}^{n_p}$
is the wavefunction coefficient matrix.

In the case of the spin-adapted DMRG, the wavefunction also has a matrix product state form. However,
the transformation matrices $\mathbf{L}^n$, $\mathbf{R}^n$ now assume a special restricted structure. In particular,
\begin{align}
\mathbf{L}^{n} &= \mathbf{U}^n \underline{\mathbf{L}} \\
\mathbf{R}^n &= \underline{\mathbf{R}} \mathbf{U}^n
\end{align}
Here $\mathbf{U}^n$ is a unitary matrix containing the Clebsch-Gordan coefficients
that construct pure spin states out of the product states $|l\rangle|n_l\rangle$, $|n_r\rangle|r\rangle$, and
$\underline{\mathbf{L}}$, $\underline{\mathbf{R}}$
are transformation matrices that map from the complete basis of pure spin states to the renormalised spin state basis.
In addition,  $\underline{\mathbf{L}}$ and
$\underline{\mathbf{R}}$ also display  a special block structure, namely states
with different spins are not mixed.
Overall, we can view the spin-adapted DMRG algorithm as carrying out an energy minimization
within the space of matrix product states, subject to the above restrictions.

\section{$3j$ coefficients}

The Clebsch-Gordan coefficients are related to the Wigner $3j$ symbols as shown below.
\begin{align}
c^{S_2 S S_1}_{m_2 m m_1}=
 (-1)^{S_2-S+m_2}
 (2S_1+1)^{1/2}\left(\begin{array}{ccc} S_2& S & S_1\\ m_2&
   m & -m_1                                              \end{array}\right) 
\end{align}
The Wigner $3j$ coefficients have
some convenient symmetry properties. Two which we  make use of
are
\begin{align}
 \left(\begin{array}{ccc} j_1 &j_2& j_3 \\ \mu_1& \mu_2 & \mu_3\\
  \end{array}\right) =& (-1)^{j_1+j_2+j_3}\left(\begin{array}{ccc} 
  j_2 &j_1& j_3 \\ \mu_2& \mu_1 &\mu_3\\
  \end{array}\right)\\
   \left(\begin{array}{ccc} j_1 &j_2& j_3 \\ \mu_1& \mu_2 & \mu_3\\
  \end{array}\right) =& (-1)^{j_1+j_2+j_3}\left(\begin{array}{ccc} 
  j_1 &j_2& j_3 \\ -\mu_1& -\mu_2 &-\mu_3\\
  \end{array}\right)
\end{align}

\subsection{Adjoint of operator} \label{sec:transpose}
The reduced matrix elements of the adjoint of a tensor operator is not the same as the adjoint of the reduced matrix elements of the tensor operator. The reduced matrix elements of the adjoint of tensor operators appearing in our spin-adapted DMRG implementation are shown below, 
\begin{align}
\langle\mu' j|| O^{0\ddag} ||\mu j\rangle =& \langle\mu j
||O^{0}||\mu' j \rangle\label{eq:ts1}\\ 
\langle\mu' j|| O^{1\ddag}||\mu j\rangle =& \langle\mu j ||O^{1}||\mu' j\rangle\\ 
\langle \mu j+1|| O^{1\ddag} || \mu' j\rangle =& (-1)\sqrt{\frac{2j+3}{2j+1}} \langle\mu' j|| O^{1} ||\mu j+1\rangle\\ 
\langle \mu j+\frac{1}{2}|| O^{1/2\ddag} || \mu' j\rangle=&\sqrt{\frac{2j+2}{2j+1}} \langle \mu' j|| O^{1/2} || \mu j+\frac{1}{2}\rangle \label{eq:ts2}
\end{align}

Here we only derive Eq. (\ref{eq:ts2}) and the other
equations can be derived in an analogous fashion. Of course one has to
remember that the adjoint here is defined as in
Eq. (\ref{eq:totran}) and in derivation below $T$ is the adjoint of $O$. In the derivation below the first equation is valid because the Clebsch-Gordan coefficient is non-zero. In fact
this Clebsch-Gordan coefficient is always equal to 1.
\begin{widetext}
\begin{align}
\langle \mu j+\frac{1}{2}|| T^{1/2} || \mu' j\rangle =& C^{j, \frac{1}{2},
  j+\frac{1}{2}}_{j, \frac{1}{2}, j+\frac{1}{2}} \langle \mu j+\frac{1}{2} j+\frac{1}{2} |T^{1/2,1/2}|
\mu' j j\rangle\nonumber\\ =& (2j+1)^{1/2} \left(\begin{array}{ccc}
  j &\frac{1}{2}& j+\frac{1}{2}\\ j& \frac{1}{2} & -j-\frac{1}{2}\\
                                                                                \end{array}\right)
 \langle\mu' j j |T^{1/2,1/2 \dag}|\mu j+\frac{1}{2}
 j+\frac{1}{2}\rangle\nonumber\\ =& (2j+2)^{1/2} (-1)^{2j+1}
 \left(\begin{array}{ccc} j+\frac{1}{2} &\frac{1}{2}& j\\ -j-\frac{1}{2}& \frac{1}{2} & j\\
                                                                            \end{array}\right) \langle\mu' j j |-O^{1/2,-1/2}|\mu j+\frac{1}{2}
 j+\frac{1}{2}\rangle\nonumber\\ =& -(2j+2)^{1/2} \left(\begin{array}{ccc}
   j+\frac{1}{2} &\frac{1}{2}& j\\ j+\frac{1}{2}& -\frac{1}{2} & -j\\
                                                                            \end{array}\right) \langle\mu' j j |O^{1/2,-1/2}|\mu j+\frac{1}{2}
 j+\frac{1}{2}\rangle\nonumber\\ =& -\sqrt{\frac{2j+2}{2j+1}} (2j+1)^{1/2}
 \left(\begin{array}{ccc} j+\frac{1}{2} &\frac{1}{2}& j\\ j+\frac{1}{2}& -\frac{1}{2} & -j\\
                                                                            \end{array}\right) \langle\mu' j j |O^{1/2,-1/2}|\mu j+\frac{1}{2}
 j+\frac{1}{2}\rangle\nonumber\\ =& -\sqrt{\frac{2j+2}{2j+1}} C^{j+\frac{1}{2},\frac{1}{2},
   j}_{j+\frac{1}{2}, -\frac{1}{2} , j} \langle\mu' j j |O^{1/2,-1/2}|\mu j+\frac{1}{2}
 j+\frac{1}{2}\rangle\nonumber\\ =&-\sqrt{\frac{2j+2}{2j+1}} \langle\mu' j
 ||O^{1/2}||\mu j+\frac{1}{2} \rangle
\end{align}
\end{widetext}
\end{document}